\documentclass[a4paper,fleqn,usenatbib]{mnras}

\usepackage[T1]{fontenc}
\usepackage{ae,aecompl}

\usepackage{graphicx}	
\usepackage{amsmath}	
\usepackage{amssymb}	

\title[Featured Spectra of FRBs]{The Integrated Pulse Profiles of Fast Radio Bursts}

\author[Q. W. Song et al.]{
Q. W. Song,$^{1,2,3}$
Y. Huang,$^{1,2}$\thanks{E-mail: huangyu@pmo.ac.cn}
H. Q. Feng,$^{4}$
L. Yang,$^{1}$
T. H. Zhou,$^{1}$
Q. Y. Luo,$^{1}$
\newauthor
T. F. Song,$^{5,6}$
X. F. Zhang,$^{5}$
Y. Liu,$^{5}$
G. L. Huang$^{1}$
\\
$^{1}$Purple Mountain Observatory, Chinese Academy of Sciences, 10 Yuanhua Road, Nanjing, 210033, China\\
$^{2}$Key Laboratory of Dark Matter and Space Astronomy, Purple Mountain Observatory, Chinese Academy of Sciences, 10 Yuanhua Road, \\Nanjing, 210033, China\\
$^{3}$Key Laboratory of Astronomical Optics $\&$ Technology, Nanjing Institute of Astronomical Optics and Technology, Chinese Academy of Sciences, \\Bancang Street, Nanjing 210042, China\\
$^{4}$College of Physics and Electronics Science, Luoyang Normal University, Luoyang, 471022, China\\
$^{5}$Yunnan Observatories, Chinese Academy of Sciences, Kunming 650011, China\\
$^{6}$College of Optoelectronic Engineering, Chongqing University, Chongqing, 400044
}

\date{Accepted XXX. Received YYY; in original form ZZZ}

\pubyear{2020}

\begin{document}
\label{firstpage}
\pagerange{\pageref{firstpage}--\pageref{lastpage}}
\maketitle

\begin{abstract}
Multi-peaked features appear on the integrated pulse profiles of fast radio burst observed below 2.5 GHz and the instantaneous spectrum of many bursts observed between 4 and 8 GHz.  The mechanism of pulse or spectrum shaping has attracted little attention. Here we show that these interference-like pulse profiles are mostly the instantaneous spectra {produced in a spiral structure} near the source region of fast radio bursts. The corresponding instantaneous spectra are coincident to the spectrum from a single electron passing through a tapered undulator. The multi-peaked spectrum observed between 4 and 8 GHz can also be explained consistently by this type of spectrum. The spectrum is invisible unless the particles in the radiation beam are bunched.  The bunching effect is probably due to the acceleration of particles in the plasma wakefield. {With a conjectured model of the wind from the source, we deduce that the radiation particles should be protons or other nuclei with energy $\gamma\sim10^4$. }
\end{abstract}




\section{Introduction}\label{sec:intro}

Fast radio bursts (hereafter FRBs) are millisecond radio flashes  of extragalactic origin. There are various proposals on their progenitors. Most proposals focus on the energy budget of the burst and try to conceive a scenario in which intense radiation can be produced. However, with  only a loose constraint on the energy budget, the discussion is divergent. There could be more and more proposals on the FRB progenitors. Moreover, without further constraints, we can never discriminate between them.

Spectrum method is often an outlet for an astrophysical process to be identified. Especially, a none single-peaked spectrum  imposes a unique constraint on the energy distribution on different frequencies. Reproducing the spectrum bring more credit to  a theory. Multi-peaked instantaneous spectra have appeared in the 4$\sim$8 GHz (the Extended Data Figure \ref{fig1} of \citealp{2018Natur.553..182M}; \citealp{2018ApJ...863....2G}) FRB observations, but have not attracted enough attention. We believe that these multi-peaked spectra are key to resolve the FRB puzzle. We will show that they are probably a type of undulator spectrum, which has been studied by scientists working on accelerators in the 1980s and known as the spectrum from a tapered undulator.

Meaning while, the pulse profiles of some fast radio bursts observed below 2.5 GHz show a multi-peaked feature, for example, the double-peaked FRB 121002 \citep{2016MNRAS.460L..30C} and many triple-peaked bursts from the repeating FRB 121102 \citep{2019ApJ...876L..23H}. For these bursts, by perturbing the dispersion measure  a little bit, we can get an oblique dynamic spectrum, in which the instantaneous spectrum  is also multi-peaked. The corresponding spectrum can also be approximated by the spectrum from a tapered undulator. The traditional dedispersion of these bursts is probably incomplete.

In this report, we present a unified description of the instantaneous spectrum of a category of FRB bursts shown none single-peaked instantaneous spectrum or integrated pulse profiles. In our theory, these FRBs are radiations from charged particles traveling in a quasi-periodic magnetic structure. A 1.4 GHz observational example is firstly presented to illustrate the basic idea. A detailed model based on the theory of radiation  from a tapered undulator (\citealp{1983PhRvA..28..319B}; \citealp{1982PhRvA..26..438S}) is then outlined. In a subsequent section, an accurate simulation to fit the observed dynamic spectrum is introduced, and a comparison is made between the simulation and the observation of two FRB bursts;   Finally, some discussions are made concerning the assumptions on the initial energy distribution of radiation particles, the low beam emittance required for the corresponding spectrum to be observed {, and the counterpart of the quasi-periodic magnetic structure near the FRB source.}

\section{Observation and theory}\label{sec:obs}
Before reaching the Earth, the radio signal of FRBs has experienced a dispersion process by traveling in cold plasma. Restoring the original signal requires correcting a time delay for each frequency channel. There is uncertainty in current dedispersion processes because the time delay or the equivalent dispersion measure (DM) is not known accurately and has to be estimated. We believe that this uncertainty is a critical issue in current FRB studies. It prevents us from recognizing a uniform pattern from FRB observation. In the rest of the paper, we will treat the dispersion measure of an FRB as a free parameter changing its recognized value as long as it is necessary.

\subsection{An observational example around 1.4 GHz}
\begin{figure*}
\centering
\includegraphics[scale=.5]{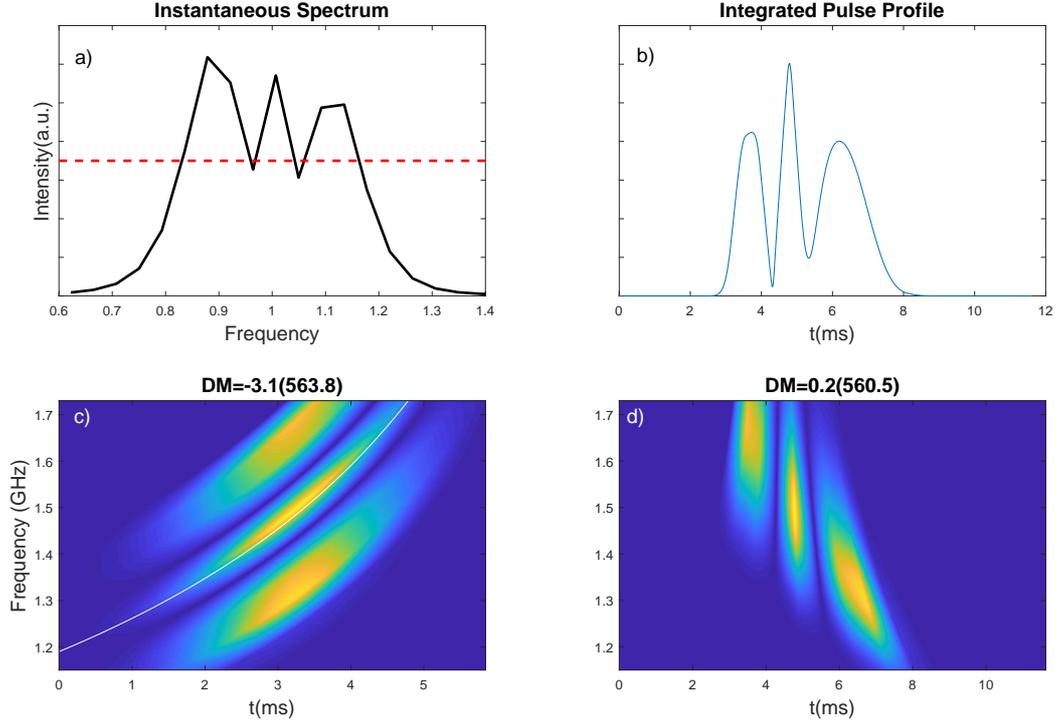}
\caption{The simulation of the dynamic spectrum of AO-06. (a) The instantaneous spectrum supposed to be observed near the source region of the burst. The dashed red line is the noise level adopted in the simulation. (b) The integrated pulse profiles of the dynamic spectrum in (d). (c) The dynamic spectrum simulated by taking (a) as the instantaneous spectrum and letting the central frequency drift along the white curve, which is a dispersive curve with dispersion measure: DM=-3.1 pc cm$^{-3}$. (d) A simulated schematic of the dynamic spectrum of AO-06 presented traditionally. Produced by dispersing the spectrum in (c) by 3.3 pc cm$^{-3}$.}
\label{fig1}
\end{figure*}

AO-06 \citep{2019ApJ...876L..23H} is a triple-peaked burst from the repeating FRB 121102 \citep{2016Natur.531..202S}. Its dynamic spectrum shows a downward drifting time-frequency structure (Fig. \ref{fig1}d). Other FRB 121102 bursts like AO-02 and GB-01 also show such a feature. Several authors have investigated the cause of the formation of the feature(\citealp{2019MNRAS.485.4091M}; \citealp{2019ApJ...876L..15W}). Their studies are based on the dynamic spectrum dedispersed traditionally, like the one presented in Figure \ref{fig1}d. However, an alternative dynamic spectrum could also be considered, in which the emission time is continuous and frequency-dependent.

We suggest that  the real implication of the structure showing a downward  time-frequency drifting is that it is ill dedispersed. The original dynamic spectrum should be the one illustrated in Figure \ref{fig1}c, which is an estimation spectrum of the burst AO-06 if it is dedispersed to 563.8 pc cm$^{-3}$, instead of 560.5 pc cm$^{-3}$. In this further dedispersed spectrum, the burst occurs in the lower frequency band first, reaches its maximum in the middle time of the burst on the middle-frequency, fades out from the higher frequency band at last. An essential feature of the newly conceived dynamic spectrum is that the instantaneous spectrum of it keeps being the same triple-peaked one throughout the radiation process. The instantaneous spectrum (Fig. \ref{fig1}a) is similar to the integrated pulse profile obtained in the usual way; both have three peaks. In this sense, we may say that the integrated pulse profiles of FRBs are the instantaneous spectra near the sources of FRBs.

\subsection{Undulator radiation and the frequency-swept dynamic spectrum}
An eligible physical interpretation of the dynamic spectrum of AO-06 (Fig. \ref{fig1}c) needs to address two observational facts: [1] the triple-peaked instantaneous spectrum; [2] the drifting of central frequency from the low frequency to the high. Undulator radiation is a known mechanism that is capable of generating both effects at the same time.

\begin{figure*}
\centering
\includegraphics[scale=.35]{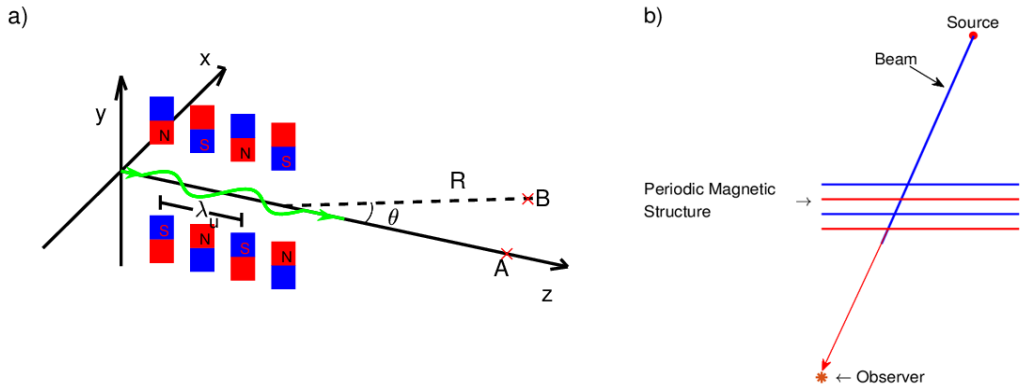}
\caption{The magnetic field causing undulated motion of charged particles. (a) The schematic of an undulator. (b) A conceived scenario of radiation process of FRBs: charged beam passing through a quasi-periodic magnetic structure at a zero angle to the observer.}
\label{fig2}
\end{figure*}

An undulator (Fig. \ref{fig2}a) is a periodic array of dipole magnets with alternating polarity. It is commonly used as an insertion device in the storage ring to produce monochromatic synchrotron radiation with relativistic particles. The frequency of radiation in the direction of angle $\theta$ is determined by equation \citep{2007psr..book.....H}:
\begin{equation}
f=\frac{2\gamma^2}{1+\frac{K^2}{2}+\gamma^2\theta^2}(\frac{c}{\lambda_u})
\label{equ1}
\end{equation}
Where $\gamma$ is Lorenz factor, $c$ is the speed of light, $\lambda_u$ is the period of the magnetic field structure. $K$ is the vector potential of the magnetic field normalized to the rest energy of the radiation particle. The radiation produced in the weak magnetic field ($K\leq1$) is the so-called undulator radiation. This equation is essentially describing a Doppler effect. The Doppler shift is angle-dependent; if you look on-axis, you get the highest frequency; if you go off-axis, the frequency gets lower and lower.

A frequency-swept dynamic spectrum can be produced by the process illustrated in Figure \ref{fig2}b, which is conceived mostly according to the circumstances between the Earth and the Sun. A beam of relativistic electrons is produced in an active source, might be a neutron star or double stars. The field in the acceleration site is strong. The radiation of AO-06 is not produced here; It is produced in a relatively weak field. After traveling {about $10^7 km$ or more}, when the ambient magnetic field get weak  enough($K\leq1$), electrons meet with a quasi-periodic magnetic structure. The radiation of the burst AO-06 is produced here.

To describe the burst AO-06, an on-axis condition (where $\theta=0$) suffices. In this case (Fig. \ref{fig2}b), the variation of $f$, the central frequency, is only determined by the variation of $\gamma$. $f$ is proportional to $\gamma^2$.  For the burst AO-06, a raw estimation gives that the low energy electrons arrived 3 ms earlier (Fig. \ref{fig1}c). With a smaller $\gamma$, they produce radiation in the lower-frequency band. The central frequency increases as the electrons arriving later have larger $\gamma$.  To offer an appreciation with magnitude, if a beam of electrons with $\gamma$ increasing from $2.07\times10^3$ to $2.37\times10^3$, sequentially passed through a periodic structure with a period of 2000 km and a weak magnetic field ($K<1$). Then the central frequency received at a distance will drift from 1.28 to 1.68 GHz similar to the one shown in Figure \ref{fig1}c.

 In addition to the central frequency drifting, the observed radiation intensity is also varying with frequency. A gaussian envelope has been used in our simulation to account for the energy distribution of radiation particles.

\subsection{Peaky spectrum from a tapered undulator}
The triple-peaked spectrum illustrated in Figure \ref{fig1}a is the squared Fourier transform of a sinusoidal function with 28 periods over which the wave vector ($k=2\pi/\lambda$) is increased by 46.5\% (hereafter the taper parameter $\eta$). This type of spectrum is known to the accelerator scientists as the radiation spectrum from a tapered undulator (\citealp{1983PhRvA..28..319B}; \citealp{1982PhRvA..26..438S}). The spectrum can be produced by passing mono-energetic electrons through a periodic magnetic structure with 28 periods over which the wave vector is increased by 46.5\%.

The forming mechanism of the spectrum can be easily apprehended through the relationship between a driven motion and driven force. For the moment, we call the radiation particles by the name of electrons, although we will show later that they are probably nuclei. The undulated motion of electrons in periodic magnetic fields is a driven motion and a harmonic oscillation in the frame of reference moving with the electrons. It can be approximated by a wave function $e^{-i\omega t}$.  {As the magnetic field is weak, the observer is kept within the radiation cone. In such a condition, the Fourier transform of the electron's apparent trajectory is  the first-order approximation of the radiation spectrum.}  The on-axis undulator radiation spectrum of a single electron is the Fourier transform of $e^{-i\omega t}$ with a Doppler shift due to the transformation between different frames.  When the spatial frequency (driven frequency) of the magnetic structure is linearly increasing, the frequency of oscillation (driven motion) also increases linearly. The motion of particles can be approximated by $e^{-i(\omega_0t+bt^2)}$, which is merely the function $e^{-i\omega t}$ with $\omega$ substituted by a linear varying term $\omega_0+bt$. The key point here is the $t^2$ term in the phasor. Because of this, the radiation spectrum, which can be approximated by the Fourier transform of this wave train, is a Fresnel integral. This process shares a similar mathematic formulism with near field (Fresnel) diffraction of visible light from narrow slits \citep{1983OptCo..45..407B}.  In optics, the triple-peaked pulse profile of AO-06 corresponds to the diffraction pattern from a slit encompassing 1.5 Fresnel zones \citep{hecht2017optics}.

From the energy perspective of view, the radiation spectrum of electrons reflects how electrons lose energy in the magnetic vector potential field through radiation. The spectrum is determined by the magnetic vector potential's spatial distribution along the traveling path of electrons. Assuming the magnetic structure to have plane polarization and the magnetic field is weak, then the spontaneous radiation spectrum is proportional to the squared Fourier transform of the transverse magnetic vector potential (the Appendix). The vector potential $A_\bot$ of a tapered magnetic field and the spectrum $S(\omega)$ measured at the radio telescope can be written as:
\begin{equation}
A_\bot=A(z)sin(k_0[z+bz^2])
\label{equ2}
\end{equation}
\begin{equation}
S(\omega)\propto|\omega\cdot Fourier(A_\bot)|^2
\label{equ3}
\end{equation}
Where $A(z)$ is the envelope of the magnetic field's amplitude along the electrons' path: $z$; it is approximated by a Gaussian function in our simulation and affects the symmetry of the spectrum; $k_0$ is the wave vector at $z=0$; $b$ is a constant and describes the linear variation of the wave vector along $z$; $\omega$ is the radiation frequency. Assuming a constant velocity along the $z$-direction, then the variable $z$ in Equation \ref{equ2} can be substituted with the time variable $t$, that is the time representation discussed in the previous paragraph.

\begin{figure}
\centering
\includegraphics[width=\columnwidth]{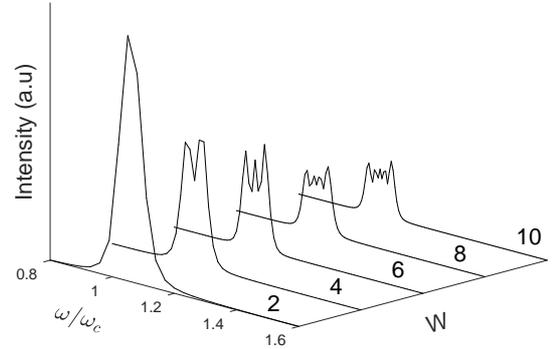}
\caption{Spectral distribution around the central radiation frequency $\omega_c$ from a tapered undulator with $\eta=0.12$.}
\label{fig3}
\end{figure}

The multi-peaked feature of the spectrum comes from the Fourier component of Equation \ref{equ3}.  A parameter $W$ can be defined as $\frac{\ln(1+\eta)\times N}{2}$ (the Appendix), where $N$ is the number of undulator periods, $\eta$ is the taper parameter.  The number of peaks appearing on a spectrum is about $W/2$ (illustrated by Figure \ref{fig3}). For example, the double-peaked FRB 121002 \citep{2016MNRAS.460L..30C} should have a $W$ value around four. Figure \ref{fig3} shows the spectrum for a general case  expressed in terms of the $W$ parameters, calculated through Equation \ref{equ3} with a symmetric envelope $A(z)$. This plot is the equivalence of Figure \ref{fig4} of the paper \citep{1988NIMPA.267..537W}.


\section{Simulation}
The theory above can explain how the peaky instantaneous spectrum are formed, which is in frequency domain only. The simulation of the observation is in a time-frequency domain. The time variation of the central frequency and average intensity of the radiation spectrum has to be assumed in our simulation.
\subsection{the procedure of simulation}
We will explain the simulation procedure through the simulation of the burst AO-06, which is a case with $W\sim5.44$. The objective of the simulation is twofold. We need to generate a dynamic spectrum whose shape correlates with the observation approximately. Meanwhile, its integrated pulse profile should be similar to the observed one. We can only strike a balance between them by adjusting eleven parameters manually iteratively. In the following, we will skip the iteration process and directly use the known value of the relevant parameters.
\begin{figure}
\centering
\includegraphics[scale=.4]{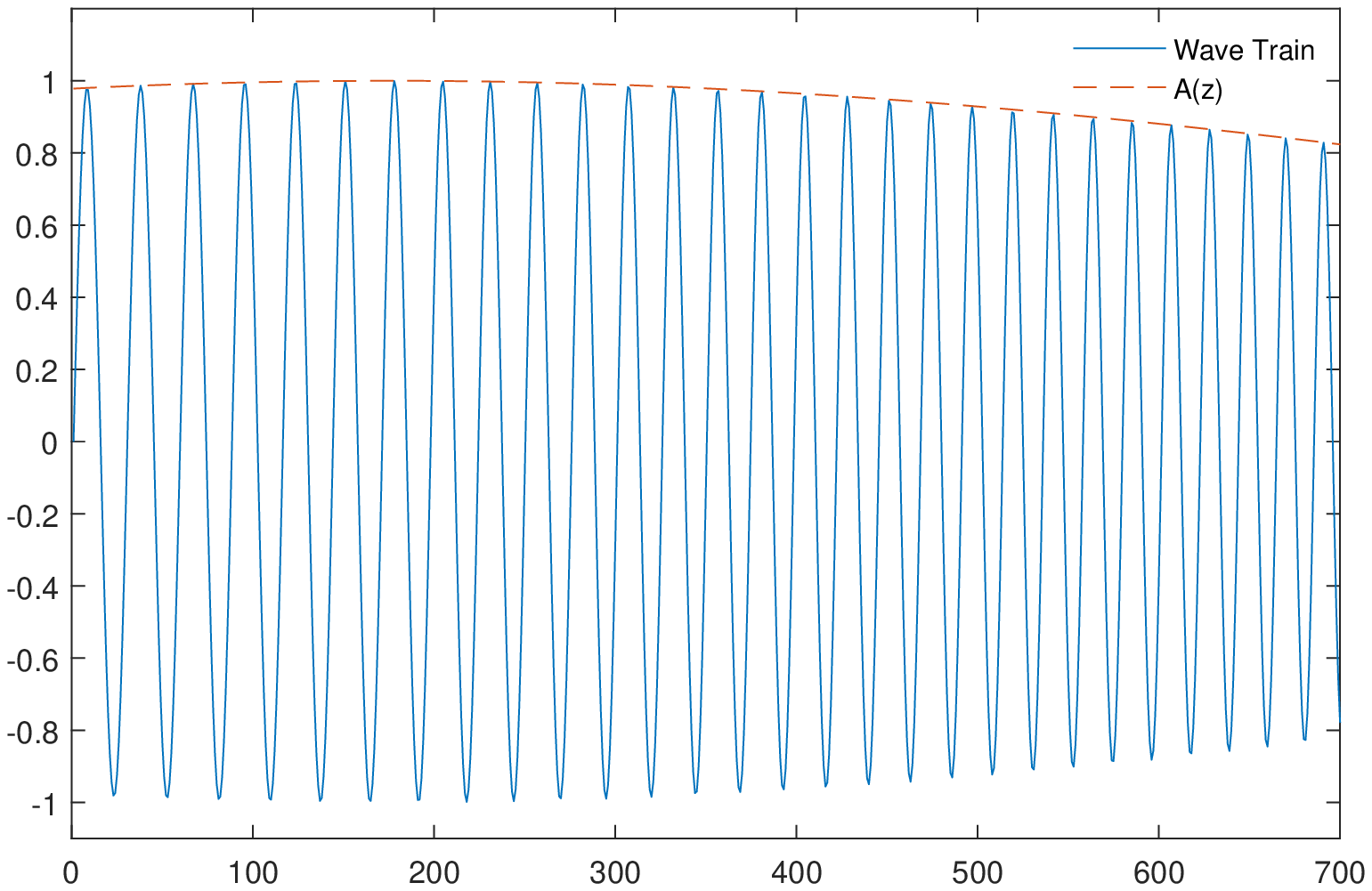}
\caption{The magnetic field breeding the burst AO-06. It is a sinusoidal function with 28.5 periods over which the wave vector increases by 46.5\%.}
\label{fig4}
\end{figure}

The main steps of the simulation are:

(1) A wave train (in Figure \ref{fig4}) is firstly generated to represent the magnetic field potential $A_\bot$ in Equation \ref{equ2}. Two parameters determine the sine component of the Equation \ref{equ2}: $W=5.44$ and $\eta=0.46$ from which the b parameter can be calculated. It is equivalent to know how many periods the magnetic structure has and how it is varying. The unit length of the undulator is defined by the periodic length at $z=0$, hence $k_0=2\pi$. The envelope $A(z)$ is approximated by a Gaussian function defined by the variable $MuWave$ and  $SigmaWave$ in the code.  { The real form of $A(z)$ maybe $\propto 1/r$ where r is the distance from the FRB source. $A(z)$ is related to the toroidal magnetic field. We'll discuss this in Section \ref{sec:dis}}.  The values of $MuWave$ and  $SigmaWave$ affect the relative heights of the peaks of the final integrated pulse profile. The length of $A(z)$ is normalized to the length of the undulator ($L\_eta$ in the code).

(2) The instantaneous spectrum in Figure \ref{fig1}a is computed according to the Equation \ref{equ3} by doing a Fourier transform to the wave train, multiplying circular frequency, and squaring itself. The red dashed line in Figure \ref{fig1}a represents the background noise level, which is a free parameter, defined by the variable $bg$ in the code.

(3) The dynamic spectrum in Figure \ref{fig1}c is created by subtracting the background noise $bg$ from the spectrum in Figure \ref{fig1}a, then moving it along the white line in Figure \ref{fig1}c, and applying a Gaussian function to modulate the radiation intensity. The white line is a dispersion curve defined by $t=4149\frac{DM}{Frequency^2}$  with the dispersion measure (hereafter $DM$) $\sim -3.1$ pc cm$^{-3}$. The value of the DM is defined by the variable $dm$ in the code. The Gaussian function is defined by the variables $sigma$ and $mu$. The unit length of the Gaussian function is defined by the frequency range of the white line. The high and low-frequency boundaries of the line determine the frequency range. They are defined by the variables $fh$ and $fl$ in the code. The output is not sensitive to $fh$ and $fl$ because we can adjust the width of the Gaussian envelope through the variable $sigma$. However, sufficient values should be assigned to them to guarantee reasonable time-frequency coverage. The parameters set in this step are corresponding to the effects related to the distribution of the radiation particles. Since the distribution is unknown, we have to guess the radiation spectrum in the time-frequency domain. The reasonability of the conjecture made here will be discussed in Section \ref{sec:dis}.

(4)	The dynamic spectrum in Figure \ref{fig1}d is obtained by dispersing the dynamic spectrum in Figure \ref{fig1}c by 3.3 pc cm$^{-3}$ further. The variable $dm1$ in the code is account for this DM lag.

A physical interpretation of the simulation above is that the original dynamic spectrum is oblique. It is produced by a beam of charged particles passing through a quasi-periodic magnetic structure. The Fourier transform of the potential of the magnetic field has three peaks. The space-time distribution of the particles results in a dispersive-like frequency drift. However, the current data processing method is incomplete; that the dynamic spectrum is presented vertically like the one in Figure \ref{fig1}d. The dynamic spectrum in Figure \ref{fig1}d is supposed to be observed in a position far away from the source region of FRB, to which the DM from the source of the burst is about 3.3 pc cm$^{-3}$.

The above approach can be easily generalized to simulate other triple-peaked bursts like the burst GB-01, AO-02, and AO-11 of FRB 121102 \citep{2019ApJ...876L..23H}, the September 17 burst of FRB 180814.J0422+73 \citep{Amiri2019} and FRB 181017 \citep{Farah2019MNRAS.488.2989F}. They are just the burst AO-06 with slightly different magnetic field configurations and the energy distributions of radiation particles. Naturally, the pulse of triple-peaked radio pulsars, like PSR J1757-2421 \citep{Yuan2017MNRAS.466.1234Y}, also can be reproduced with the same approach. The simulation codes of the burst AO-06, GB-01, AO-02, AO-11, and AO-13 written in Matlab language and the Matlab code generating all the figures in this paper can be downloaded online\footnote{The Matlab codes can be downloaded from: \url{https://github.com/huangyu7991/frb.git}}.  The output pictures of the codes are also available online for none Matlab user's references.

The simulation shows that two important parameters can be estimated from the observation for sure. One is N, the number of period of the quasi-period structure; the other one is $\eta$, the variation speed of the period length. Besides, two parameters of the envelope of the magnetic field can also be estimated, supposing the envelope is of Gaussian shape. The relative bandwidth of the burst and the shape of the spectrum constrain these parameters.  {From Equation 1, $\lambda_u=2\gamma^{2}\lambda$ (where $\lambda$ is the observed wavelength,  and assuming $\theta=0, K\approx0$), with the precise knowledge of  $\gamma$ or $\lambda_u$, the other one can be calculated.}

\subsection{Simulation of a 4-8 GHz burst}
The instantaneous spectra of many 4$\sim$8 GHz bursts of FRB 121102  (\citealp{2018Natur.553..182M}, \citealp{2018ApJ...863....2G}) are almost already multi-peaked. The dynamic spectrum of these high-frequency bursts can be reproduced with the approach introduced above. The difference is that the  central frequency of  the  high-frequency instantaneous spectrum is not drifting. So we only need a function to mimic the variation of radiation intensity and need not to fit the slope of the dispersive curve.

\begin{figure*}
\centering
\includegraphics[scale=.35]{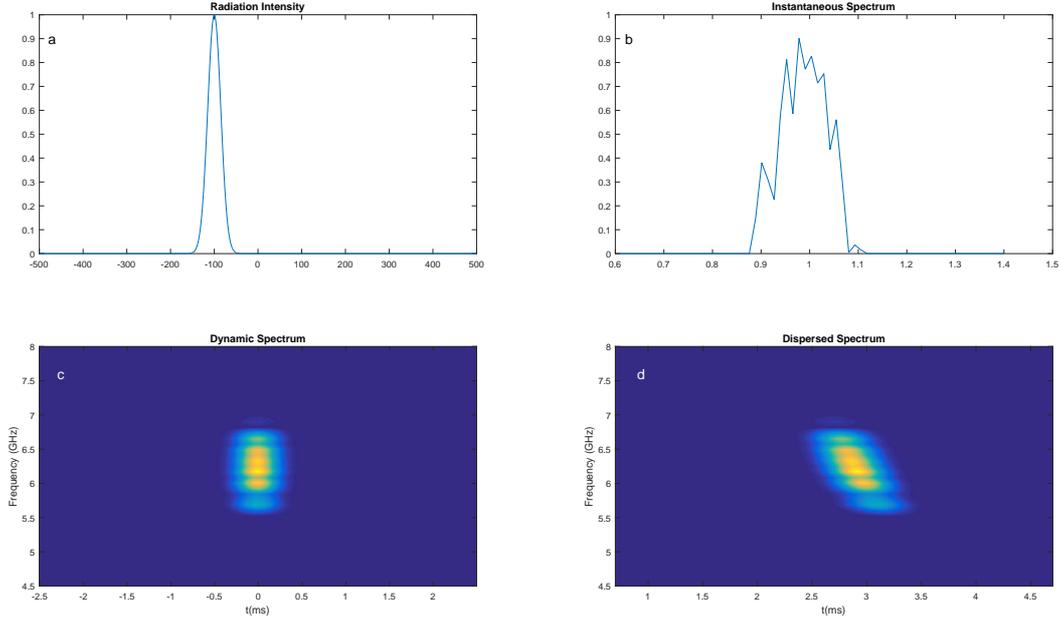}
\caption{The simulation of the burst 11O. (a) Assumed  {time} variation of radiation intensity. (b) The spectrum of a wave train with W=12.5. (c) The dynamic spectrum obtained by let the spectral intensity of b vary according to the function in the a. (d) The dynamic spectrum obtained by dispersing c with DM=9.3 pc cm$^{-3}$}
\label{fig5}
\end{figure*}

The burst 11O  of FRB 121102 \citep{2018ApJ...863....2G} has six apparent peaks on its instantaneous spectrum. Its simulation result is presented in Figure \ref{fig5}d. To produce this pattern, we first create the instantaneous spectrum in Figure \ref{fig5}b by doing a Fourier transform to a sinusoidal wave train with N=90 and $\eta=32\%$. Then
let the  spectral strength vary with time as the Gaussian function shown in Figure  \ref{fig5}a, we obtain the dynamic spectrum shown in Figure \ref{fig5}c. Dispersing this spectrum  with DM=9.3 pc cm$^{-3}$, we get Figure \ref{fig5}d.

The interpretation of the simulation is: the burst 11O is produced by mono-energetic particles moving through a periodic magnetic structure with 90 periods over which the wave  {vector} increased by 32$\%$. The number of the radiation particles or the coherence of the beam varies with time as the curve in Figure \ref{fig5}a. The original energy distribution in the time-frequency domain looks like the pattern in Figure \ref{fig5}c. However, the burst in the publication is similar to the one shown in Figure \ref{fig5}d, for using a DM which is 9.3 pc cm$^{-3}$ smaller than the real one.

The Gaussian function used in Figure \ref{fig5}a is not ideal because it is symmetric while the observed time variation of the radiation intensity  is not. Nevertheless,  for simplicity, we pertain to the symmetric case. It is enough to constrain the energy in the certain time-frequency domain and illustrate the basic idea. The other issue at odds with the observation is that the spectrum in Figure \ref{fig5}b has seven peaks, while the observation shows six  apparent peaks. There must be a peak in the observation at the noise level and it is not resolved. To keep the  different patches in Figure \ref{fig5}c separating from each other as the observation, we have no freedom to use parameters otherwise.

\subsection{Comparison with the observation}
\begin{figure}
\centering
\includegraphics[width=\columnwidth]{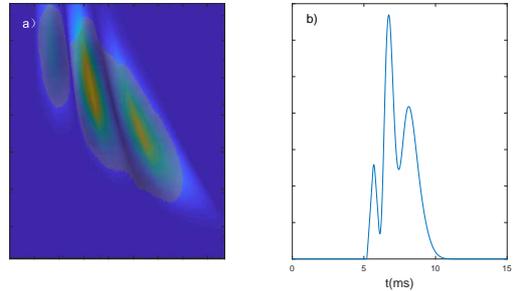}
\caption{A comparison between the simulation and the observation of the burst AO-02. a) the simulated dynamic spectrum ( blue and yellow) of the burst AO-02, blended with the observation (gray). b) the integrated pulse of the dynamic spectrum in a).}
\label{fig6}
\end{figure}

 The burst AO-02 is another triple-peaked burst of FRB 121102 \citep{2019ApJ...876L..23H}. A comparison between the simulation and the observation of the burst AO-02 is in Figure \ref{fig6}. Its simulation can be obtained by dispersing a DM$\sim-4.5$ pc cm$^{-3}$ dynamic spectrum by 5.5 pc cm$^{-3}$. The blue and yellow image in Figure \ref{fig6}a is the simulated time-frequency structure. The gray-colored part of the image is created from a 3D printable data\footnote{The 3D printable data is downloaded from: \url{https://www.thingiverse.com/thing:2723399/files}} of the burst.  It can be seen that the two parts correlate to each other very well, except that the simulated separation between the two strips on the right-side does not match that of the observation. By overlapping the two dynamic spectra each other, we are not doing a strict mathematical fitting; we are showing that the  process proposed by us can generate an  energy  distribution in time-frequency domain similar to that of an FRB.  The process can give a constrained energy distribution only, just better than a pure theory full of formulas.

 In the simulation of AO-06, we cannot keep the strip in the top-left corner of Figure \ref{fig1}d strictly vertical as the observation. Otherwise, the envelope of the other two strips cannot match the observation very well.

 The discrepancies between the simulation and observation may stem from two main factors: (1) Our simulation can only reflect the radiation process but not the whole process, including propagation and data collection. (2) The slope of the original dynamic spectrum can not be described by a dispersion relationship strictly as what we have assumed.

\begin{figure}
\centering
\includegraphics[scale=.35]{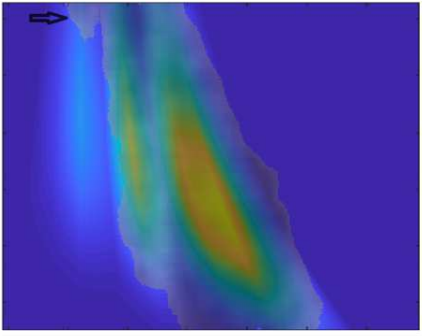}
\caption{A picture with the observation (gray colored) and the simulation(blue-yellow colored) of the burst AO-13 blended.}
\label{fig7}
\end{figure}

A comparison between the simulation and the observation of the burst AO-13 \citep{2019ApJ...876L..23H} brings an inconspicuous feature into the attention. Although the pulse of the burst has two peaks, it is probably a triple-peaked burst with the high-frequency part buried under the background noise level. The fact is revealed by the residual radiation in the high-frequency range, marked by the black arrow in Figure \ref{fig7}. A small amount of observed radiation(gray colored) locates in the position where the third strip is supposed to appear. We intentionally lower the background noise level of the image to let the leftmost strip appeared.

\section{Discussion}\label{sec:dis}
This is the first time that undulator radiation is suggested to be in an astrophysical process. Lots of issues can be questioned. We will limit the discussion on critical assumptions or implications.

(1) The spectrum discussed here is essentially a single-particle spectrum. An implicit assumption in our work is that the spectrum from an electron can approximate the radiation spectrum of the electron beam at any instance.  Usually, the assumption is not valid, and the spectrum is not visible. Because  particles in an ordinary beam generally are traveling  at different   velocities and at an angle to each other, which means different $\theta$ and $\gamma$ of Equation \ref{equ1}. Hence, the central frequencies of the radiation spectrum from different particles are different.  The multi-peaked spectrum will be smooth-off in the observed radiation, which is a summation of the radiation from all particles.

Only under certain circumstance, the multi-peaked spectrum is visible. That is when the particles in the electron beam are highly directional and mono-energetic. One possibility for this happening is that the beam is bunched. Hence, particles in the same bunch travel at the same velocity in the same direction. A known mechanism generating the bunching effect is the plasma wakefield acceleration (PWFA) process, which happens when a relativistic charged beam moves through plasmas \citep{Chen1985PhRvL..54..693C}. The PWFA process is self-modulated (\citealp{Kumar2010PhRvL.104y5003K}; \citealp{Pukhov2011PhRvL.107n5003P}). The beam will split into a series of bunches. This effect may also be an answer to the question of why the radiation of FRBs is so intense. The bunching effect will increase the radiation strength by order of the electron number in a bunch.

(2) Some FRB properties inferred here can be associated with the behaviors of radio pulsars, especially the Crab pulsar. The Crab pulsar is known to have a highly variable DM. While the average value of its DM is 56.8 pc cm$^{-3}$, the variation of the DM is about 50 pc cm$^{-3}$(\citealp{2003ApJ...596.1142C}). This behavior can be explained by an oblique dynamic spectrum, whose slope is variable from pulse to pulse. The intensity of some giant pulses(GP) from the Crab pulsar can exceed 2MJy (\citealp{2007ApJ...670..693H}). Such an increase in the radiation intensity by orders can also be a manifestation of the bunching effect.

\begin{figure}
\centering
\includegraphics[width=\columnwidth]{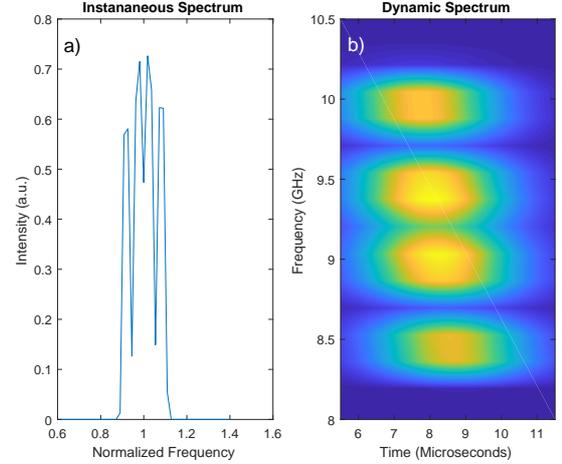}
\caption{The simulation of a GP. a) The spectrum derived from a sinusoidal function with N=36.6, $\eta=30\%$. b) The simulated dynamic spectrum of the GP.}
\label{fig8}
\end{figure}

This study supports the view that the giant pulses of the Crab pulsar and FRBs share the same radiation mechanism. As \citet{2018ApJ...863....2G} have notified that,  similar to the FRB bursts reported by them, the instantaneous spectrum of some GP (\citealp{2007ApJ...670..693H}) is also multi-peaked. The dynamic spectra of GP can also be reproduced with our method and have a similar explanation.  Figure \ref{fig8} is a reproduction of the dynamic spectrum of a GP (Figure \ref{fig6} of the \citealp{2007ApJ...670..693H}). Again for simplicity, we use a symmetric Gaussian function to mimic the time variation of the radiation intensity, although the time variation in the observation is asymmetric.

(3)	A significant assumption made in our simulation of low-frequency radiation is that the energy distribution of the radiation particles will result in a dispersive-like frequency drift, as shown by the white line in Figure \ref{fig1}c. The rationality of such an assumption comes from the selection effect that has been inevitably introduced into the FRB observation through the current FRB searching strategy. In practice, the FRB searching algorithm dedisperse time series to find a single pulse. Such kind of algorithm cannot recognize any burst deviating much from a dispersive-like frequency drift. The searching strategy itself might be the reason why there are FRBs non-repeatable. It is because the initial time-frequency drift is not always dispersive-like.

A drifting central frequency is nothing new for radio transients. In the dynamic spectrum of a GP(Figure \ref{fig7} of the \citealp{2007ApJ...670..693H}), the central frequency of the spectrum is increased by about 500 MHz within $6\mu s$. An explanation for such drifting is that the  energy of the radiation particles ($\gamma$ in Equation 1) is increasing.


(4) { What is the undulator-like magnetic field in practice? One possibility is that it is a spiral structure formed around the source of FRBs. The varying periods correspond to the different pitches of the spiral.}

 {Spiral structures are often seen in the Universe.  The Parker spiral (\citealp{1958ApJ...128..664P}) of our solar system is one close to us. The solar magnetic field  can not keep closed configuration at any distance. Far away from the Sun, when the magnetic field gets weak enough,  the plasma of the Sun will blow up. A current sheet will  form. The magnetic field on the two sides of the sheet point in the opposite direction, which leads to an electric current $J=\bigtriangledown\times B\neq 0$. Because of the solar rotation, the current sheet rolls up, forming a spiral.}

\begin{figure}
\centering
\includegraphics[scale=.38]{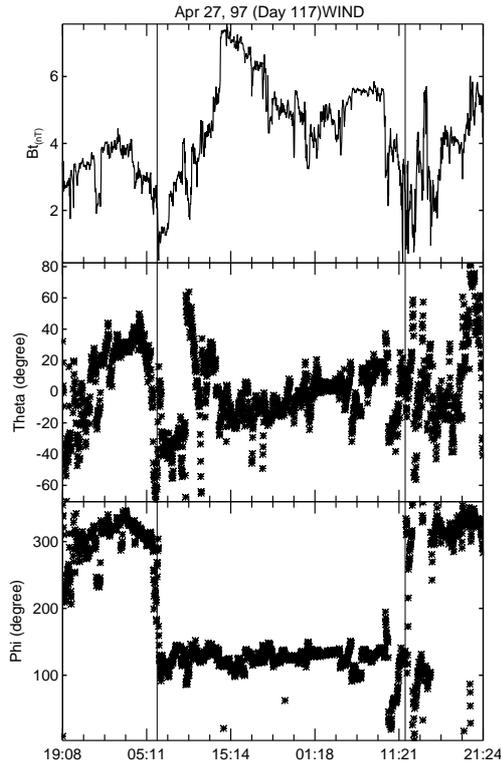}
\caption{The magnetic field observed by the WIND satellite in the solar wind near the Earth. The upper panel is the time variation of the total magnetic field in nT. The middle panel shows the elevation angle of the magnetic field, and the lower panel shows the azimuthal direction of the magnetic field in the ecliptic plane.}

\label{fig9}
\end{figure}

\begin{figure}
\centering
\includegraphics[scale=.5]{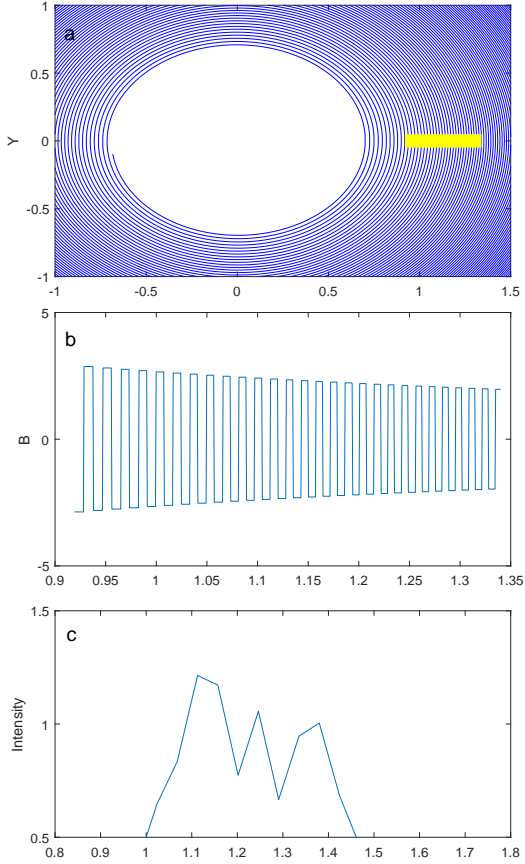}
\caption{ (a)The conceived spiral. The unit of X and Y coordinate is  $10^{7} km$. The yellow bar marks the trajectory of radiation particles.  (b) The azimuthal component of the magnetic field  (unit $10^{-8}$ T) along the trajectory (unit $10^7 km$).   (c) The normalized spatial spectrum of the wave train in (b). }
\label{fig10}
\end{figure}

 {An observation example of the magnetic field around the current sheet  in the solar wind  is plotted in Figure \ref{fig9}. It is observed by the WIND satellite (\citealp{1995SSRv...71..207L}) near the Earth between  the 27th and 29th of 1997 April.  Note the azimuthal direction of the magnetic field in the lower panel; it is about $300^{\circ}$ at first, then about $100^{\circ}$, then $300^{\circ}$ again, which implies the satellite has crossed two current sheet structures. If the satellite can keep crossing current sheet structures, we may expect the magnetic field will vary periodically like the one shown in Figure \ref{fig10}.b.  Such an alternating toroidal field is proposed to exist in the rotational equator of a pulsar (\citealp{1990ApJ...349..538C}) or any oblique rotators. The magnetic field on each side of the current sheet is planar. Linearly polarized FRBs, like FRB 150807 (\citealp{2016Sci...354.1249R}) should have been bred in such type of magnetic field. In the middle panel of Figure \ref{fig9}, it is the magnetic field in the poloidal direction. We believe that the magnetic field in this direction is also essential. It provides a drifting field and deflects the radiation cone from the observer. Otherwise,  particles will keep traveling in the observer direction and generate a radiation spectrum with too many peaks. The poloidal component can not be too small if the spectrum only has several peaks.}

 {As the exact mathematical form of the spiral is unknown,  we conceive a spiral  to get some quantitative understanding of the problem. After all, the key point here is the varying pitches which the most spirals have. We choose to generate a Fermat's spiral because we find that the spectrum derived from this type of spiral is closed to the observation.  The spiral is described by the equation:}
\begin{equation}
\phi=\frac{r^2}{r_L\times b}
\end{equation}

 {Where $\phi$ is the azimuthal angle; $r$ is the distance from the source; b=0.046 AU is a parameter of Parker's spiral(\citealp{1958ApJ...128..664P}), which marks the boundary of a steady state of a magnetohydrodynamic system; $r_L$ is the radius of the light cylinder of a pulsar. We assume the source is a pulsar and normalize the distance by the geometric average of the two parameters. Other parameters adopted are  rotation period 1.7 second, and the total magnetic field is $\sim1.6\times10^{-3} T$ at 47700 km (\citealp{2016JPlPh..82e6302P}). The azimuthal component of the magnetic field is proportional to $1/r$ for $r\gg b$ and equal to zero for $r\leq b$. This treatment is  similar to that of Parker's. Such a theoretically expected relation $B\propto1/r$ (\citealp{1958ApJ...128..664P}, \citealp{2013LRSP...10....5O}) is approximately consistent with the simulated envelope $A(z)$ in Figure \ref{fig4}. }

 {The generated spiral is plotted in  Figure \ref{fig10}.a. The yellow bar marks the expected trajectory of the radiation particles. The path crosses 28 pitches of the spiral,  over which $\lambda_U$ decrease from $1.9\times10^5 km$ to $1.3\times10^5 km$ ($\eta=45\%$).  The variation of the azimuthal component of the magnetic field along the trajectory  is plotted in Figure \ref{fig10}.b. Its Fourier transform (Figure \ref{fig10}.c)   has three peaks and the relative bandwidth is close to the one plotted in Figure \ref{fig1}.a. If a beam of charged particles, whose energy $\gamma$  increase from $1.8\times10^4$ to  $2.1\times10^4$ from  beginning to end, pass the trajectory, a dynamic spectrum  similar to the one in Figure \ref{fig1}.c will be produced. Note the periodic function here is not necessarily sinusoidal. It can be  periodic intermittence or any other periodic function.}

 {In the simulated scenario, the radiation particles should be protons or other nuclei; they can not be electrons. Undulator radiation is produced in a place where $K \leq 1$. Here, $K$ is  412 for electrons, and $\sim0.2$  for protons. The parameter $K$ is given for electrons by:}
\begin{equation}
K=\frac{eB_0\lambda_{0}}{2\pi m_0 c^2}=93.4(\frac{B_0}{Tesla} )(\frac{\lambda_0}{Meter})
\end{equation}

 { Though the spiral above is  conceived purely mathematically, the order of the simulated periodic length $\lambda_U$ is reasonable. If the wind speed is $ \sim0.5c$ (the escape velocity of a neutron star), and the rotation period is $\sim 1s$, the order of $\lambda_U$ could be $\sim10^5 km$.  The spiral is identical to the orbit of an electron in a cyclotron, where the electron energy $E=\frac{1}{2}mv^2$  increases by an equal amount  each turn, i.e., $E\propto\phi$; the angular velocity $\omega=v/r$ is a constant $\sim eB/m$, hence $r\propto v\propto\sqrt{E}\propto \sqrt{\phi}$}.

 { We have also done  the simulation  with Parker's spiral. The  results of the two simulations are generally agreed. In both simulations, the radiation is produced at a distance $\sim10^7km$ from the source by protons or more massive nuclei with energy $\gamma\sim10^4$ in a magnetic structure with $\lambda_U\sim10^5km$. With Parker's spiral, a triple-peaked spectrum can also be obtained. Still, the relative positions of different peaks do not fit the observation very well, which is related to the specific non-linear $\lambda_U$ variation. Matlab codes of both simulations are available online.}

\section{Summary}
According to our theory, two critical factors are behind FRBs: a quasi-periodic magnetic field and a bunching effect due to the relativistically moving of charged particles through plasmas. The former one is responsible for the featured instantaneous spectrum; the latter one is indispensable for the visibility of the spectrum and is an account for the extremely high radiation intensity.  {The periodic magnetic structure is likely a spiral formed around FRB sources, and the radiation particles are probably protons or other nuclei. }

The stimulation of this study is the similarity between a one-dimensional fresnel curve and the integrated pulse profile of some low-frequency FRBs. In this paper, we build a mathematical connection between them through the mechanism of undulator radiation. We have shown that the description of these low-frequency bursts applies to the high-frequency phenomena equally. The multi-peaked feature is essentially the spatial spectrum of a periodic magnetic structure near the source region of FRBs. Mathematically, the spectrum is the Fourier transform of a periodic function with increasing periodic length.

\section*{Acknowledgments}
This research is supported by the Opening Project of Key Laboratory of Astronomical Optics \& Technology, Nanjing Institute of Astronomical Optics \& Technology, Chinese Academy of Sciences. It is also supported by NSFC 11533009 U1631135, U1931138 and U1731241. Author contributions: Song, Qiwu and Huang, Yu did the theoretical and numeric work collaboratively. Other authors contributed to the investigation of possible mechanisms and the properties of the magnetic field in the interplanetary space. We thank NASA CDAWeb for making WIND magnetic field data available.

%

\appendix
\section{The Radiation Spectrum as the Spatial Spectrum of Magnetic Field.}
The content here is a physical interpretation of the formula in Section 2 of \citet{1982PhRvA..26..438S} in an on-axis case. The focus is a fundamental physical understanding, but not math strictness. The vector potential of a plane-polarized magnetic field inside the undulator with a linear taper can be written as:
\begin{equation}
\overrightarrow{A}_\bot(z)=A(z)\cos[\int_0^zk(z)dz]\overrightarrow{x}=A(z)\cos([1+\frac{\eta}{2L(\eta)}z]k_0z)\overrightarrow{x}
\end{equation}
Where $\eta$ is the taper parameter, $L(\eta)$ is the undulator's length, $k_0$ is the undulator's wave vector at $z=0$. $A(z)$ is the amplitude. A similar representation can be found in Equation (1.2) of \citet{1983PhRvA..28..319B}. It can be seen that the $b$ term in Equation \ref{equ2} is $\frac{\eta}{2L(\eta)}$.

Far away from the undulator, the radiation spectrum of a relativistic electron is given as follows \citep{2007psr..book.....H}:
\begin{equation}
\frac{dW}{d\Omega d\omega}=\frac{e^2\omega^2}{4\pi^2c}|\int_0^L\widehat{n}\times[\widehat{n}\times\overrightarrow{\beta}(z)]e^{\frac{i\omega}{c}\int_0^Z[1-\widehat{n}\cdot\overrightarrow{\beta}(z)]dz}dz|^2
\label{equa2}
\end{equation}
Where $\widehat{n}$ is the direction of observation, $L$ is the length of the undulator, $\overrightarrow{\beta}(z)$ is the electron velocity at position z normalized to light speed in a vacuum.

Consider the on-axis case where $\widehat{n}=\widehat{z}$, and ignore the small difference between the longitudinal velocity $\beta_z$ and $\beta$, then the phasor can be simplified as:
\begin{equation}
i[\frac{\omega}{c}\int_0^Z[1-\beta_z(z)]dz]=i[\frac{k}{2\gamma^2}z]
\end{equation}
The spectrum of Equation \ref{equa2} can be expressed in this form:
\begin{equation}
\frac{dW}{d\Omega d\omega}=\frac{e^2}{4\pi^2c}\cdot\omega^2\cdot[\int_0^L\beta_\bot(z)e^{i[\frac{k}{2\gamma^2}z]}dz]^2
\label{equa4}
\end{equation}
It can be seen that the spectral shape can be approximated by the Fourier transform of the driven motion $\beta_\bot(z)$. The $2\gamma^2$ in the denominator is according to the Doppler shift. The $\omega^2$ term will cause asymmetry to the spectrum. For narrow band spectrum, it can be neglected.
\begin{equation}
\beta_\bot(z)=-\frac{eA_\bot(z)}{\gamma mc^2}
\label{equa5}
\end{equation}
Inserting the above expressing of $\beta_\bot(z)$ into Equation \ref{equa4}, we may see that the radiation spectrum represents the spatial spectrum of the magnetic potential along the electron's path. The complete form of the spectrum is:
\begin{equation}
\frac{dW}{d\Omega d\omega}=\frac{e^3}{4\pi^2\gamma mc^3}\cdot\omega^2\cdot|\int_0^LA(z)\cos([1+\frac{\eta}{2L(\eta)}z]k_0z)e^{i[\frac{k}{2\gamma^2}z]}dz|^2
\label{equa6}
\end{equation}

\section{The definition of the parameter W}
This parameter is equivalent to the near field parameter defined by \citet{1988NIMPA.267..537W}, where 2$\sqrt{W}$ can be thought of dimensionless slit width.

Combining Equation (13) of \citet{1988NIMPA.267..537W}, Equation (1.4) of \citet{1983PhRvA..28..319B}:
\begin{eqnarray}
\alpha L^2=4\pi W\\
L\approx N\lambda_0 \frac{\ln(1+\eta)}{\eta}
\end{eqnarray}
Where $L$ is the length of the undulator, $N$ is the number of undulator periods, $\frac{\alpha}{2}$ is the coefficient before the $z$ square term in Equation \ref{equ2} (hence $\alpha=\frac{2\pi}{\lambda_0 L} \eta)$. We may get:
\begin{eqnarray}
W=\frac{\alpha L^2}{4\pi}=\frac{N\ln(1+\eta)}{2}
\end{eqnarray}
The physical understanding of W is that it is a measure of the peak number of a spectrum, how many peaks will appear on a spectrum is determined by how long(represented by the parameter N) is the magnetic structure and how fast($\eta$) is the period varying. The  peak number of a spectrum is about W/2.

%








%
%


\bsp	
\label{lastpage}
\end{document}